\documentclass[12pt,preprint]{aastex}

\begin{document}

\title{\large \bf Gravitational Microlensing Influence on X-ray Radiation from Accretion Disk of Active Galaxies}

\author{Predrag Jovanovi\' c}
\email{pjovanovic@aob.bg.ac.yu}

\affil{Astronomical Observatory, Volgina 7, 11160 Belgrade, Serbia and Montenegro}







\begin{abstract}
Gravitational microlensing is a very useful tool for investigating the
innermost part of lensed quasars, especially for studying a
relativistic accretion disk around a massive black hole (BH)
supposed to exist in a quasar's center. Here we present a short overview of our
recent investigations in this field.
\end{abstract}

\keywords{accretion, accretion disks --- line: profiles --- galaxies: active --- gravitational lensing --- X-rays: galaxies}

The aim of the dissertation was to investigate the influence of
gravitational microlensing on detected X-ray radiation from
accretion disks of active galactic nuclei (AGNs). For this
purpose, the focus was set to the analysis of Fe K$\alpha$
spectral line and X-ray continuum variations due to gravitational
microlensing. Besides, a significant part of the dissertation was
devoted to the optical depth of microlensing (i.e. to the
probability to observe these effects) and to the corresponding time
scales.

The initial assumption was the existence of a super massive black
hole ($10^7 - 10^9\ M_\odot$), surrounded by an accretion disk that radiates in
X-rays, in the center of all types of AGNs. Accretion disks could
have different forms, dimensions and emission, depending on the
type of central BH, being rotating (Kerr metric) or non-rotating
(Schwarzschild metric). Except for effects due to disk
instability, its emission could also be affected by gravitational
microlensing, especially in the case of the gravitationally lensed
quasars \citep{Chart02a,Dai03,Chart04}.

\begin{figure}
\includegraphics[width=5.25cm]{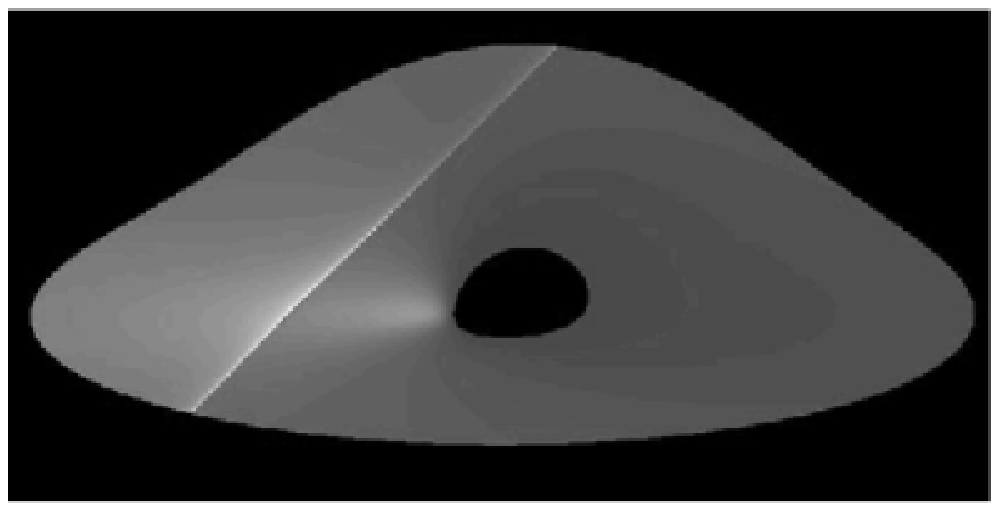}
\hfill
\includegraphics[width=10.75cm]{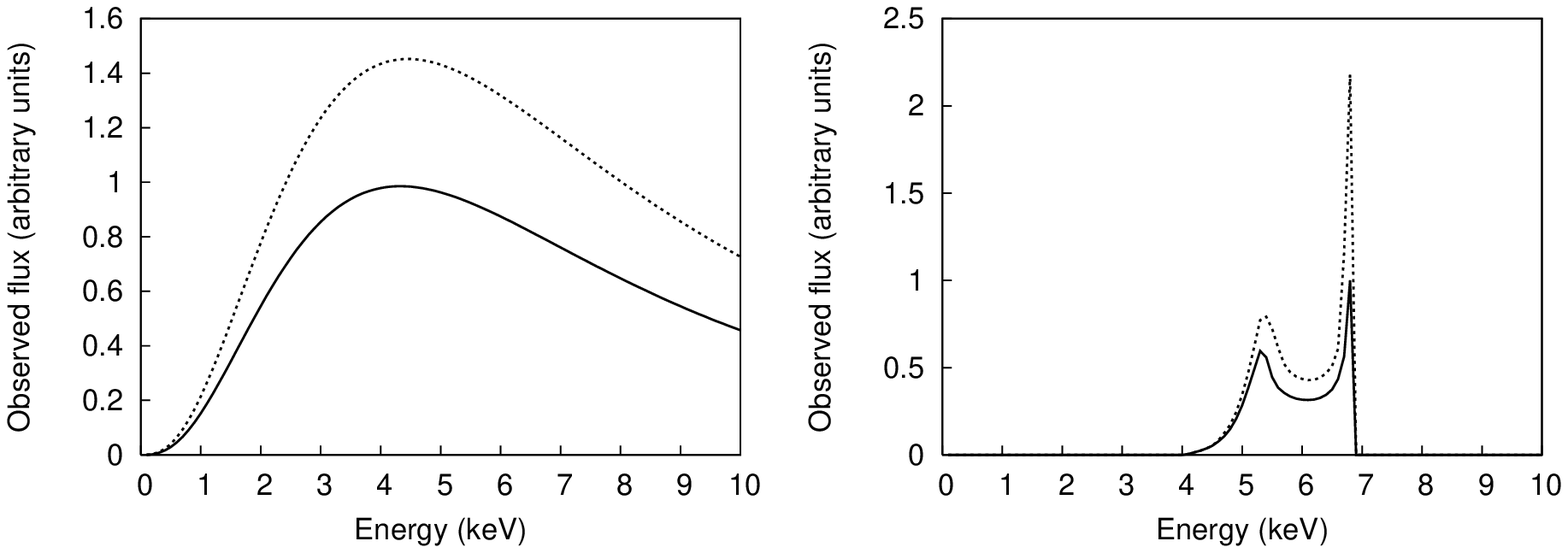}
\caption{Illustration of the straight-fold caustic crossing over an accretion disk (left),
the corresponding deformations of the X-ray continuum (middle) and the Fe K$\alpha$ line (right).
Undeformed profiles are presented by solid and deformed by dotted lines.\label{fig1}}
\end{figure}

The disk emission was analyzed by numerical simulations, based on a
ray-tracing method in a Kerr metric, taking into account only photon
trajectories reaching the observer's sky plane (see \citet{lcp03b} and references therein).
The influence of microlensing on a standard accretion disk was studied using three types of a microlensing model:
point-like microlens, straight-fold caustic and quadruple microlens (microlensing
pattern). An illustration of the straight-fold caustic crossing over an
accretion disk in the Kerr metric and the corresponding effects on the shapes of the
X-ray continuum and the Fe K$\alpha$ line are shown in Fig. \ref{fig1}.

\begin{figure}
\centering
\includegraphics[width=0.75\textwidth]{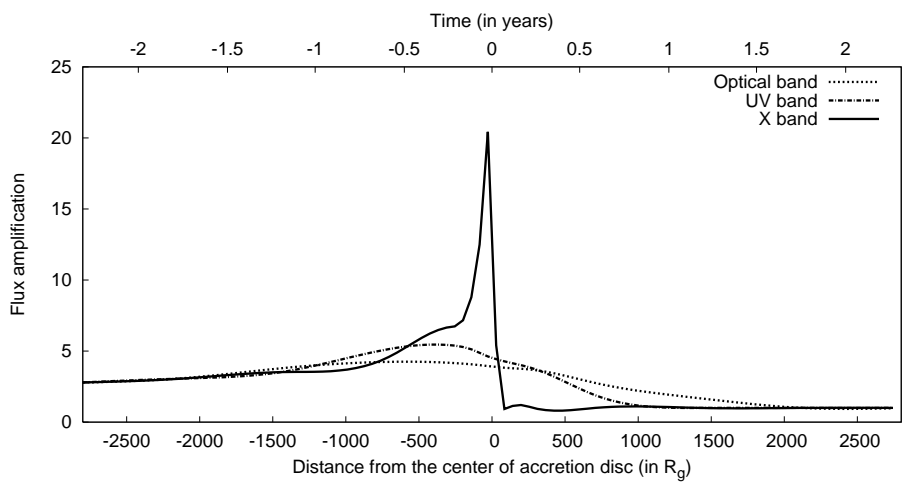}
\caption{The variations of normalized total continuum flux in
optical (3500-7000 \AA), UV (1000-3500 \AA) and X (1.24-12.4 \AA\ i.e. 1-10 KeV) band due to microlensing by a caustic crossing over accretion disk
in Schwarzschild metric.
Time scale corresponds to the QSO 2237+0305 gravitational lens, where redshifts of microlens and source are
$z_l=0.04$ and $z_d=1.69$, respectively \citep{pj06}.
\label{fig2}}
\end{figure}

Some conclusions of the dissertation are:
\begin{enumerate}
\item Gravitational microlensing can produce significant variations
and amplifications of the line and continuum fluxes. These deformations of the X-ray radiation depend on
both the disk and microlens parameters. For more details see \citet{lcp01,lcp03a}
\item Even very small mass objects could produce noticeable changes
in the Fe K$\alpha$ line profile during a microlensing event. This is because of the small dimensions of the
X-ray emitting region \citep{lcp03b,lcp02}
\item Microlensing can satisfactorily explain the excess in the iron line emission observed in three gravitational lens
systems: MG J0414+0534 \citep{Chart02a},  QSO 2237+0305 \citep{Dai03} and H 1413+117 \citep{Chart04}. More details can
be found in \citet{pj03,pj05a,pj05b} and \citet{lcp03b,lcp06}
\item On the basis of these investigations, one can expect that the Fe K$\alpha$ line and X-ray continuum amplification
due to microlensing can be significantly larger than the corresponding effects on optical and UV emission lines and continua
(see Fig. \ref{fig2} and \citet{lcp02a,lcp02b,lcp03b,pj06})
\item The optical depth for microlensing of X-ray emitting region by compact objects from the bulge/halo of the host galaxy is
very small, $\tau\sim10^{-4}$ \citep{lcp02a,lcp03b,afz04}
\item The optical depth for gravitational microlensing by cosmologically distributed deflectors could be significant and could
reach $\tau\sim0.01-0.1$. In that case, the
maximum optical depth ($\tau\sim0.1$) could be expected if dark matter forms cosmologically distributed
compact objects (see \citet{afz04,afz05})
\item Although microlensing is an achromatic effect, it can induce wavelength dependent variations of the X-ray continuum of
$\sim30\%$ due to the radial distribution of temperature in an accretion disk \citep{lcp06}
\item Analysis of three different microlensing time scales (see Fig. \ref{fig2}) led to the conclusion that the variations of the X-ray continuum
are the fastest
($\sim$ several months) in comparison to the variations of the optical and UV continua, which are weaker and much slower ($\sim$ several
years) \citep{pj06}
\end{enumerate}

According to the results of the dissertation, one can conclude that monitoring of gravitational lenses in the X-ray
spectral range may help us to
understand the physics and to reveal the structure of the innermost parts of active galaxies, particularly their
relativistic accretion disks.

\acknowledgments

This work is a part of the project: "Astrophysical Spectroscopy of
Extragalactic Objects" supported by the Ministry of Science, Technologies
and Development of Serbia.


\clearpage

\clearpage

\end{document}